# Severe violation of the Wiedemann-Franz law in quantum oscillations of NbP

Pardeep Kumar Tanwar [1], Md Shahin Alam [1], Mujeeb Ahmad [1], Dariusz Kaczorowski [2] and Marcin Matusiak [1,2,*]

1. *International Research Centre MagTop, Institute of Physics, Polish Academy of Sciences, Aleja Lotnikow 32/46, PL-02668 Warsaw, Poland*
2. *Institute of Low Temperature and Structure Research, Polish Academy of Sciences, ul. Okólna 2, 50-422 Wrocław, Poland*

The thermal conductivity ($\kappa$) of the Weyl semimetal NbP was studied with the thermal gradient and magnetic field applied parallel to [0 0 1] direction. At low temperatures $\kappa(B)$ exhibits large quantum oscillations with frequencies matching two of several determined from the Shubnikov – de Haas effect measured on the same sample with analogous electrical current and magnetic field orientation. Both frequencies found in $\kappa(B)$ originate from the electron pocket enclosing a pair of Weyl nodes. The amplitude of the oscillatory component of the thermal conductivity turns out to be two orders of magnitude larger than the corresponding value calculated from the electrical conductivity using the Wiedemann – Franz law. Analysis of possible sources of this discrepancy indicates the chiral zero sound effect as a potential cause of its appearance.

* Corresponding author, email: matusiak@magtop.ifpan.edu.pl

**Introduction**

Topologically non-trivial materials exhibit a variety of extraordinary transport phenomena. This applies to the charge transport, where, for example, coexistence of conductive surface and insulating bulk in three-dimensional topological insulators [1] or negative longitudinal magnetoresistance in topological semimetals [2] are expected. In addition, there are many intriguing effects associated with the heat transport in such materials. For instance, the presence of the thermal chiral anomaly in $Bi_{1-x}Sb_x$ [3] or a mixed axial–gravitational anomaly in NbP [4] were reported.

Among the novel phenomena exclusive to the topological materials is the chiral zero sound (CZS) effect [5], named after the zero sound effect which may appear in the Fermi liquid due to the oscillating deformation of the Fermi surface [6]. Alternatively, CZS can occur in semimetals with multiple pairs of Weyl nodes, and its signatures are supposed to be visible in the specific heat and thermal conductivity [5]. In the latter, CZS provides an additional channel for chargeless heat transfer, which leads to violation of the Wiedemann – Franz law that assumes a fixed ratio between the charge and heat fluxes. Recently, the unusual magnetic field dependence of the thermal conductivity in semimetallic TaAs was attributed to the emergence of the chiral zero sound [7].

Here, we present our studies on the thermal transport properties in another Weyl semimetal, namely NbP. The analysis of the experimental results indicates that the thermal conductivity of NbP at low temperatures is also governed by the CZS effect.

**Methods**

Single crystals of NbP were grown by chemical vapor transport method using iodine as a transporting agent, as described in the literature [8]. The crystals had typical dimensions 2 x 2 x 1 mm$^3$, metallic luster and were stable in air. Their chemical composition was checked by energy-dispersive X-ray analysis using a FEI scanning electron microscope equipped with an EDAX Genesis XM4 spectrometer. The result confirmed homogeneous single-phase material with stoichiometry nearly equal to the nominal one. Single crystal X-ray diffraction (XRD) experiment was performed on a Kuma-Diffraction four-circle diffractometer equipped with a CCD camera, using Mo-Kα radiation. The results confirmed the tetragonal nonsymmorphic space group I4$_1$md (No. 109) with the lattice parameters very close to those reported in the literature [8].

For the transport measurements, a rectangular bar with dimensions 1.9 x 0.45 x 0.4 mm$^3$ was cut from a suitable single crystal with the longest side of the sample oriented along the [0 0 1] direction (crystallographic c-axis).

The electrical resistivity was measured using a four-point technique with alternating electric current flowing along [0 0 1]. The experiments were carried out in the temperature range 1.7 – 300 K and in magnetic fields up to 14.5 T applied parallel to the electric current. For thermal conductivity measurements, the isolated heater method was used with a Micro-Measurements strain gauge as a heater and Cernox chip as a base thermometer. Thermal gradient, applied along [0 0 1], was measured with a 25 μm constantan-chromel thermocouple, calibrated in magnetic field, which was attached to the sample through a few mm long and 100 μm thick silver wires. An additional thermocouple was mounted between the base and the heater to calculate the actual sample temperature. The whole setup was suspended on the 10 μm Kevlar threads and covered with a silver radiation shield. Magnetic field was applied parallel to the thermal gradient. The DC technique was used for field sweeps, where the field varied between -14.5 and +14.5 T to extract the field-symmetric component of the signal. The quasi-AC mode was employed for temperature ramps.

**Results**

Figure 1 presents the magnetic field ($B$) dependence of the electrical resistivity ($\rho$) of NbP measured at $T = 1.7$ K with electrical current applied along the c–axis, oriented parallel to $B$. At this temperature, the resistivity initially rises steeply with increasing field, but above $B \approx 4$ T, this trend changes and $\rho$ slowly decreases. The negative magnetoresistance we observe might be due to the chiral magnetic anomaly (CMA) emerging in Weyl semimetals [8-10]. Its presence suggests that the intravalley scattering time in NbP is shorter than the intervalley scattering time [8]. Another characteristic feature of $\rho(B)$ is the appearance of pronounced oscillations, which can be attributed to the Shubnikov - de Haas (SdH) effect. The field dependencies of the electrical resistance of NbP measured at higher temperatures are shown in Fig. S1 of the Supplemental Material (SM). As can be inferred from that data, CMA vanishes above $T \approx 50$ K along with the smearing of the SdH effect.

The left-hand inset to Fig. 1 shows the oscillatory component of $\rho(B)$ ($\Delta\rho(B)$), at $T = 1.7$ K in the field range 2.5 – 12 T. This was chosen to avoid the deviation from periodicity observed in the $\rho(B)$ above $B \sim 12$ T, where one can expect an anomalous behavior from low-frequency oscillations reaching their quantum limit [11]. The oscillatory component was extracted from the total resistivity by subtracting a polynomial, which is a standard method of removing a non-oscillatory background. The low order (3$^{rd}$) function was used to limit the risk of unintentionally removing low frequency oscillations. The right-hand inset to Fig. 1 presents the fast Fourier transform (FFT) calculated from $\Delta\rho(B)$, where several peaks appear: 4.9 ($F_1$), 9.1 ($F_2$), 16 ($F_3$), 29 ($F_4$) and 40 ($F_5$) T. All of them were theoretically predicted to appear in NbP for $B \parallel c$ [12] and already observed in other experiments [8,12]. The small hump at 70 T is likely an artificial result of not perfectly subtracted background in the high field limit (see Fig. S3 in SM), although a similar frequency was also observed by Z. Wang et al. [8].

The band-structure calculations revealed two electron pockets (E1 and E2) and two hole pockets (H1 and H2) to be present in NbP at the Fermi level [12]. The $F_1$ and $F_2$ oscillations are supposed to originate from the extremal orbits in the electron pocket E2, $F_3$ and $F_4$ represent the cross-sections of the electron pocket E1, while $F_5$ comes from the hole pocket H1. The other hole pocket H2 does not contribute to quantum oscillations when $B$ is oriented along the c-axis. It's important to note that the theory predicted as many as 12 pairs of Weyl nodes of type I and II present in the electronic structure of NbP [13-15].

Figure 2 presents the temperature dependence of the thermal conductivity ($\kappa$) of NbP. The $\kappa(T)$ data is dominated by a large maximum near 40 K, observed also in the previous study of the compound [16]. As displayed in the inset of Fig. 2, in the low temperature limit, the $\kappa(T)$ variation can be well approximated with the function $\kappa(T) = aT + bT^3$. In a simple approach, the fitting parameters are related to the electronic ($a$ coefficient) and phononic ($b$ coefficient) contributions to the thermal conductivity. The cubic term describes the phonon thermal conductivity in the boundary scattering region for crystals with perfect long-range order, but for imperfect crystals it is expected to dominate over a wider range of temperature [17]. The inset in Fig. 2 presents also the electronic thermal conductivity ($\kappa_{WF}$) calculated from the electrical conductivity ($\sigma = \rho^{-1}$, since there is no Hall signal for $j \parallel B$) using the Wiedemann – Franz (WF) law:

$$L = \frac{\kappa}{\sigma T}\left(\frac{e}{k_B}\right)^2, \qquad (1)$$

where $L$ is the Lorenz number (for the Fermi liquid $L$ equals to $\pi^2/3$ in the regime of elastic scattering of conduction electrons), $e$ stands for the elementary charge, and $k_B$ is the Boltzmann constant. As can be inferred from the figure, the value of $\kappa_{WF}/T$ turns out to be nearly temperature independent below $T \approx 20$ K with $\kappa_{WF}/T \approx 0.17 \pm 0.04$ W m$^{-1}$ K$^{-2}$, where the error is based on the estimated uncertainty in determination of the geometrical factor. Notably, this value matches well the parameter $a$, derived from the low temperature $\kappa(T)$ data (i.e. $a = 0.44 \pm 0.22$ W m$^{-1}$ K$^{-2}$), which is the electronic thermal conductivity divided by $T$.

The magnetic field dependencies of the thermal conductivity of NbP measured at a few different temperatures in the configuration $B \parallel c \parallel \nabla T$ are presented in Fig. 3. The most prominent feature of $\kappa(B)$ taken at $T \leq 7.8$ K is the presence of oscillations which are visible in fields $B > 6$ T. They were found periodic with $B^{-1}$ and the inset to Fig. 3 shows the fast Fourier transforms calculated for the 6 – 14.5 T field range after a linear background was subtracted from the raw $\kappa(B)$ data. In FFT, two main frequencies are visible ($F_2 = 15$ T and $F_3 = 30$ T) where the latter coincides with the frequency observed previously for NbP in the $\kappa(B)$ oscillations measured for $B \parallel c \perp \nabla T$ [16]. As emphasised in Ref. 17, the amplitude of oscillations in $\kappa(B)$ was much larger than the values derived from the WF law.

**Discussion**

The magnetic field dependencies of the low-temperature thermal conductivity in NbP resemble in many aspects those reported for TaAs [7]. The main similarity is the presence of giant magnetic quantum oscillations in $\kappa(B)$, which are clearly in antiphase with oscillations in $\kappa_{WF}(B)$, as exemplified in Fig. 4 for $T \approx 5$ K. Similar behaviour of $\kappa(B)$ and $\kappa_{WF}(B)$ was found at other temperatures – see Fig. S2 of the Supplemental Material. Remarkably, the oscillatory component in $\kappa(B)$ of NbP is approximately 100 times larger than that in the corresponding $\kappa_{WF}(B)$ function. The previous study suggested the chiral zero sound as a possible origin of thermal quantum oscillations of this magnitude [7]. Nevertheless, several other explanations also need to be considered.

The first scenario to discuss is that the oscillations observed in $\kappa(B)$ of NbP are due to the contribution from the electronic system which massively violates the WF law. In fact, there were theoretical and experimental works indicating that in topologically non-trivial materials the Lorenz number can be enhanced. For example, when $B \parallel \nabla T$ this may happen due to the thermal chiral anomaly [3,18,19]. However, in such a case the expected enhancement of $L$ is much smaller than that observed in NbP. Moreover, our experimental data suggests that the electronic system obeys the WF law in weak magnetic fields. As shown in Fig. 4, the initial drop of the thermal conductivity when the magnetic field departures from zero is well matched by a drop in the electronic contribution resulting from large magnetoresistance. Additionally, a strong indication that the observed $\kappa(B)$ variation is not a straightforward result of changes in the electronic thermal conductivity is related to the fact that in strong magnetic field the oscillations in $\kappa_{\text{WF}}(B)$ anti-correlate with those in $\kappa(B)$.

Since NbP is a semimetal with multiple electron and hole pockets, a role of ambipolar contribution to the thermal conductivity ($\kappa_{eh}$) needs to be considered. This is associated with the presence of electron – hole pairs [17] and can be expressed as [20, 21]:

$$\kappa_{eh} = \left(\frac{\pi^2 k_B}{3e}\right)^2 T \left(\frac{\sigma_e \sigma_h}{\sigma_e + \sigma_h}\right) \left(\frac{k_B T}{E_{F,e}} + \frac{k_B T}{E_{F,h}}\right)^2, \qquad (2)$$

where $E_F$ stands for the Fermi energy, and indexes $e$ and $h$ denote the electronic and hole parameters, respectively. Equation 2 implies that $\kappa_{eh}$ becomes small, if the electron and hole electrical conductivities are not compensated. Another limiting factor is the temperature that needs to be comparable to the Fermi temperatures of the electron and hole pockets – for example in graphene the ambipolar contribution is substantial above $T \approx 50$ K but the WF law recovers in the low temperature limit [22].

In case of NbP the negative sign of the Hall coefficient at low temperature and high magnetic field suggests that the electronic transport in the range of studies is dominated by electrons [23]. Measurements performed for $j \parallel c$ and $B \parallel a$ presented in Fig. S5 support this conclusion. However, even if one assumes the perfect electron-phonon compensation ($\sigma_e = \sigma_h = \sigma/2$), the ambipolar thermal conductivity calculated at $T = 5$ K remains very small. Namely, for the Fermi energies taken from band structure calculations: $E_{F,e} = 57$ meV and $E_{F,h}$ = 5 meV [13] the resulting $\kappa_{eh} \approx$ 8 x 10$^{-3}$ W m$^{-1}$ K$^{-1}$, which is more than two orders of magnitude smaller than the amplitude of oscillations in $\kappa(B)$.

Furthermore it is worth to mention that an alignment of electron and hole bands in the momentum space is required, since a distance between the pockets will result in a long recombination time, and thus suppression of the ambipolar conduction [24]. Based on these remarks, one can conclude that the ambipolar effect can unlikely be responsible for the $\kappa(B)$ behaviour observed in NbP.

The third possible origin of the distinct oscillations in $\kappa(B)$ of NbP that needs to be considered is magnetic field influence on the phonon thermal conductivity ($\kappa_{ph}$), which seems to dominate the zero-field heat transport at temperatures covered in our experiment. Although the motion of phonons, as chargeless quasiparticles without magnetic moment, should not be affected by magnetic field, there are exceptions to this rule [25-27]. On the other hand, the frequency of oscillations in $\kappa(B)$, matching those from the Shubnikov – de Haas effect, suggests some relation between these two phenomena. An imaginable connection may lie in an important role of electrons in phonon scattering. In such a case, the electronic density of states oscillating with field can influence the thermal transport of phonons via alternation of their mean-free path.

This mechanism was identified as an origin of large quantum oscillation in $\kappa(B)$ of elemental antimony, where phonons are indeed strongly coupled to electrons [28]. Noteworthy, in Sb, the oscillations in $\kappa(B)$ and $\kappa_{WF}(B)$ are out of phase, like revealed for TaAs and NbP. However, the very large amplitude of the oscillations in $\kappa(B)$ would imply that electrons dominate the phonon scattering at low temperatures. At odds with this scenario, predicting a $\kappa(T) \sim T^2$ dependence in the low temperature limit [29,30], the experimental data of NbP follows the function $\kappa(T) = aT + bT^3$ (cf. the inset to Fig. 2). Such a dependence requires that both the electron and phonon mean free paths are $T$ – independent below $T \sim 20$

K. In this temperature range, $\rho$ changes only by about 6% (see Fig. S1) suggesting almost constant the mean-free path of electrons. The phonon mean-free path ($\lambda$) can be calculated using the specific heat data [16,31] and resulting $\lambda \approx 4$ μm at $T = 5$ K suggests that the limiting factor at this temperature is scattering of crystallographic defects. To roughly estimate the temperature range in which the *T*-independent scattering dominates we used, as an approximation, $\lambda(T)$ reported for solid helium [32]. This allowed us to estimate that $\lambda$ should be nearly constant below $T \approx 22$ K that is observed in the experiment.

Here, it is worth mentioning that recent ultrasonic studies of NbP revealed rather modest magnitude of the relative amplitude of oscillations of the phonon velocity (*v*), measured with the propagation vector of acoustic wave and the magnetic field direction being both parallel to c-axis [33]. The obtained value $\Delta v/v \approx 10^{-4}$ is similar to the result reported for TaAs [7]. Another hint that phonons in NbP are not dominantly scattered by electrons comes from the thermoelectric power (*S*) data. In the presence of strong electron-phonon coupling, one may expect a large contribution from the phonon-drag effect, which affects *S* approximately by a factor $\frac{\tau_p}{\tau_{pe}+\tau_p}$, where $\tau_{pe}$ is the phonon-electron relaxation time, and $\tau_p$ represents the phonon relaxation time for all the other phonon scattering processes [34]. If $\tau_{pe}$ were comparable or shorter than $\tau_p$, the phonon drag thermopower would be significant. Quite the opposite, the low temperature $S(T)$ dependence of NbP was judged to be unlikely influenced by phonon-drag [16].

Considering all the above arguments, one is left with a recently proposed mechanism, limited only to the Weyl semimetals, i.e. explanation of the presence of giant magnetic quantum oscillations in the thermal conductivity in terms of the chiral zero sound effect [5,7]. The CZS phenomenon can occur in a semimetal with multiple pairs of Weyl nodes, and consists in the formation of collective "breathing modes" of Fermi surfaces in different Weyl cones, which can be treated as neutral bosonic excitations of Weyl fermions. Modes originating from different Weyl points have opposite phase shifts, which leads to cancellation of charge currents, but important contribution to the heat transport. The CZS velocity was predicted to be inversely proportional to the electronic density of states at the Fermi level, hence the oscillations in $\kappa(B)$ caused by this effect are supposed to appear in anti-phase with the oscillations in $\kappa_{WF}(B)$ [7].

The occurrence of giant oscillations in $\kappa(B)$ that are out-of-phase to those in $\kappa_{WF}(B)$ is not the only signature of the CZS effect in NbP. One should note that the non-oscillatory component of $\kappa(B)$ originating from CZS is expected to rise linearly with $B$ in the weak field and low temperature region [5]. Our results shows in fact such a field dependence of the thermal conductivity at the lowest temperature, which at $T = 7.8$ K seems to be masked by the opposite behaviour of the phononic contribution clearly visible at $T = 19$ K. Figure 4 shows also the field dependence of the thermal conductivity in the CZS scenario calculated within the model introduced by Z. Song et al. [5] (see SM). Taking into account approximations used in the model, the calculated curve matches the experimental data fairly well. The resulting Debye temperature for the CZS mode : $\Theta_{CZS}(B = 14.5 \text{ T}) \approx 55$ K, and the ratio between the intervalley and intravalley scattering times at $T = 5.2$ K: $\tau_0 / \tau_{CZS} \approx 4.5 \times 10^{-4}$ placing NbP in the chiral limit [5].

**Summary**

We investigated the low-temperature thermal transport in the Weyl semimetal NbP. The $\kappa(B)$ dependencies, obtained for $B \parallel c \parallel \nabla T$, showed pronounced quantum oscillations with an amplitude about two orders of magnitude larger than could be expected from the Wiedemann – Franz law and being in antiphase with $\kappa_{WF}(B)$. We also observed that the non-oscillatory component of $\kappa(B)$ increases with field at low temperature. The critical discussion of possible sources of the observed features led to the conclusion that they unlikely originate from the electronic or ambipolar thermal conductivities, and they are also not caused by magnetic field driven modulation of the phonon – electron scattering. We suggest that the highly anomalous behaviour of $\kappa(B)$ established for NbP emerges due to the chiral zero sound effect, which can be considered as evidence for the presence of multiple Weyl states in this compound. We believe that the CZS effect may become a new and efficient tool in the study of topologically non-trivial materials.

**Acknowledgments**

This work was supported by the Foundation for Polish Science through the IRA Programme co-financed by EU within SG.

**Competing financial interests:**

The authors declare no competing financial interests.

**Data availability**

All of the relevant data that support the findings of this study are available from the corresponding author upon reasonable request.

**M.Matusiak ORCID iD**: 0000-0003-4480-9373

**Figures**

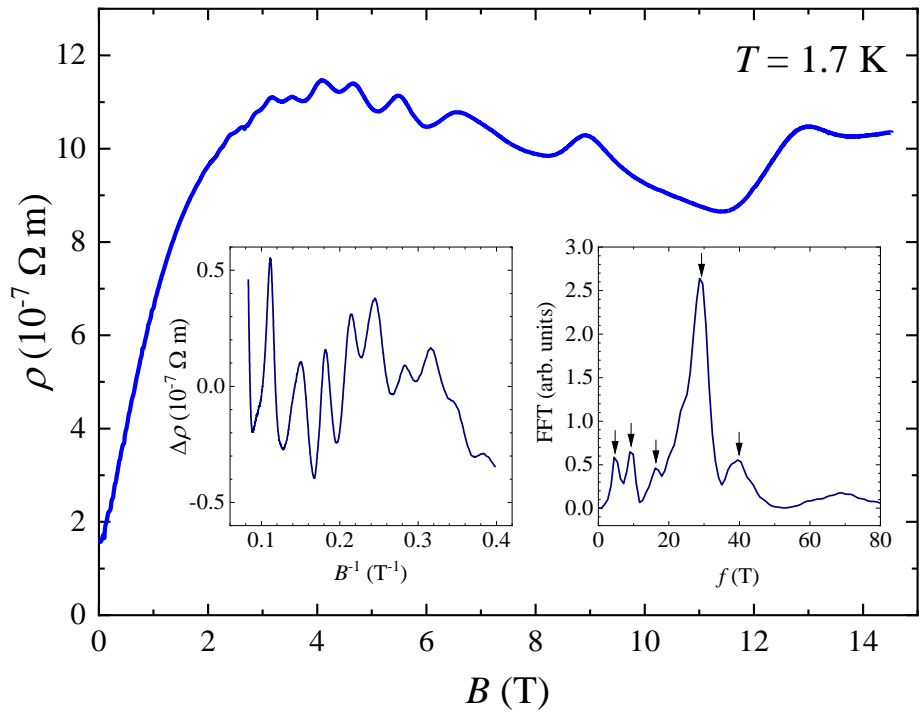

**Figure 1.** (Color online) Magnetic field dependence of the resistivity of NbP at $T = 1.7$ K measured along c-axis and the magnetic field parallel to the current. The left-hand inset shows oscillatory part of the resistivity (after subtraction of 3rd order polynomial background) plotted versus inversed field in the range 2.5 – 12 T. The right-hand inset shows the fast Fourier transform of this signal with arrows denoting peaks related to: 4.9, 9.1, 16, 29 and 40 T frequencies.

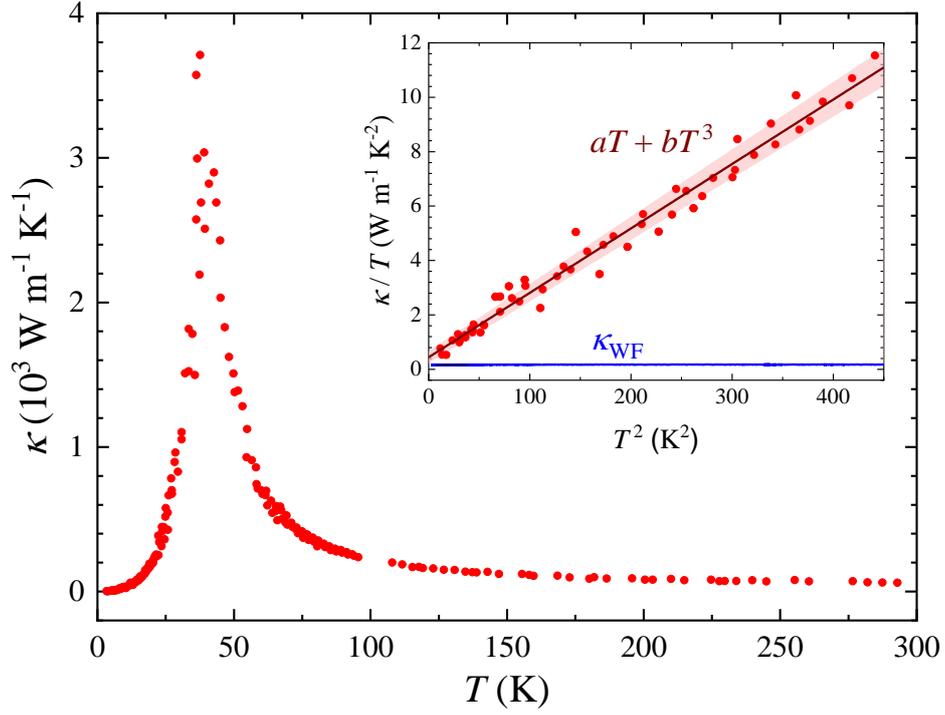

**Figure 2.** (Color online) Temperature dependence of the zero-field thermal conductivity in NbP. Inset shows the low temperature part ($T \leq 22$ K) of the $\kappa(T)$ dependence with the dashed line representing $\kappa(T) = aT + bT^3$ fit, where light-red region denotes uncertainty of the measurement. Solid blue line is shows the electronic thermal conductivity calculated from the Wiedemann – Franz law.

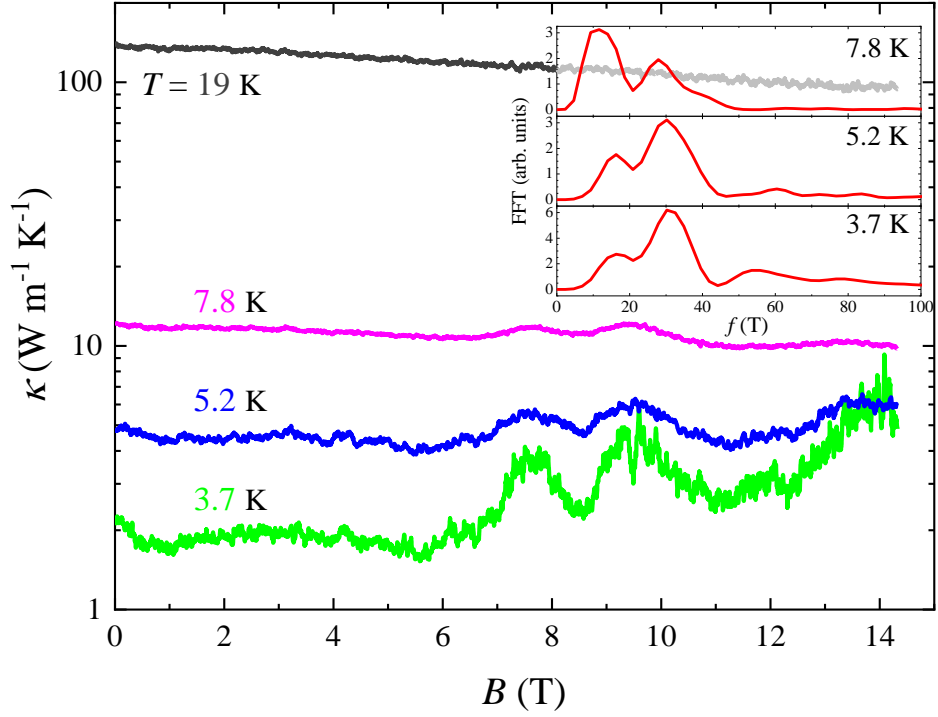

**Figure 3.** (Color online) Magnetic field dependence of the thermal conductivity ($B \parallel c \parallel \nabla T$) of NbP for several temperatures plotted in the semi-logarithmic scale. Inset presents the fast Fourier transform of the oscillatory component of $\kappa(B)$ for $T$ = 3.7, 5.2, 7.8 K (from the bottom up) showing two frequencies: ~ 15 and 30 T.

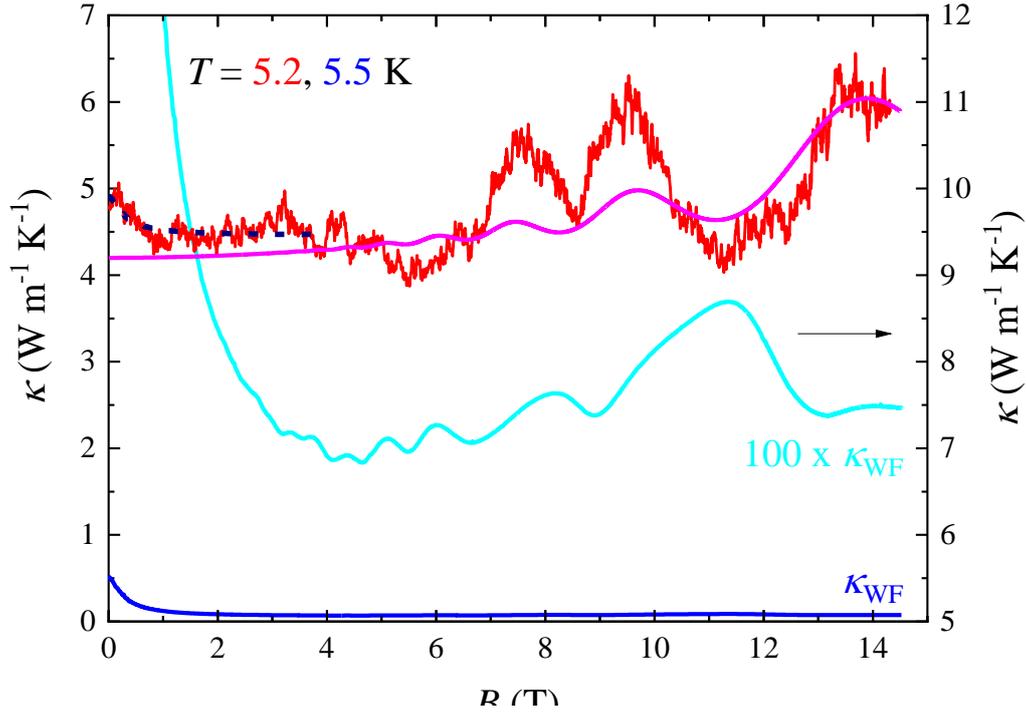

**Figure 4.** (Color online) Magnetic field dependence of the thermal conductivity of NbP at $T = 5.2$ K (uppermost red line) compared to the electronic thermal conductivity calculated from the Wiedemann – Franz law at $T = 5.5$ K (lowermost blue line). The dashed line is part of the $\kappa_{WF}(B)$ dependence shifted vertically to match $\kappa_{WF}$ and $\kappa$ values in the zero field limit. The middle (cyan) line shows $\kappa_{WF}(B)$ multiplied by 100 (right axis). The topmost (pink) line is the total thermal conductivity including the CZS contribution calculated from the model introduced in Ref. [5].

**References**

[1] L. Fu, C. L. Kane, and E. J. Mele, *Topological Insulators in Three Dimensions*, Phys. Rev. Lett. **98**, 106803 (2007). DOI: https://doi.org/10.1103/PhysRevLett.98.106803

[2] D. T. Son and B. Z. Spivak, *Chiral Anomaly and Classical Negative Magnetoresistance of Weyl Metals*, Phys. Rev. B **88**, 104412 (2013). DOI: https://doi.org/10.1103/PhysRevB.88.104412

[3] D. Vu, W. Zhang, C. Şahin, M. E. Flatté, N. Trivedi, and J. P. Heremans, *Thermal Chiral Anomaly in the Magnetic-Field-Induced Ideal Weyl Phase of $Bi_{1-x}Sb_x$*, Nat. Mater. **20**, 1525 (2021). DOI: https://doi.org/10.1038/s41563-021-00983-8

[4] J. Gooth, A. C. Niemann, T. Meng, A. G. Grushin, K. Landsteiner, B. Gotsmann, F. Menges, M. Schmidt, C. Shekhar, V. Süß, R. Hühne, B. Rellinghaus, C. Felser, B. Yan, and K. Nielsch, *Experimental Signatures of the Mixed Axial-Gravitational Anomaly in the Weyl Semimetal NbP*, Nature **547**, 324 (2017). DOI: 10.1038/nature23005

[5] Z. Song and X. Dai, *Hear the Sound of Weyl Fermions*, Phys. Rev. X **9**, 021053 (2019). DOI: https://doi.org/10.1103/PhysRevX.9.021053

[6] L. D. Landau and E. M. Lifshits, *Course of Theoretical Physics* (Pergamon, Oxford, 1980), Vol. 5.

[7] J. Xiang, S. Hu, Z. Song, M. Lv, J. Zhang, L. Zhao, W. Li, Z. Chen, S. Zhang, J.-T. Wang, Y.-F. Yang, X. Dai, F. Steglich, G. Chen, and P. Sun, *Giant Magnetic Quantum Oscillations in the Thermal Conductivity of TaAs: Indications of Chiral Zero Sound*, Phys. Rev. X **9**, 031036 (2019). DOI: 10.1103/PhysRevX.9.031036

[8] Z. Wang, Y. Zheng, Z. Shen, Y. Lu, H. Fang, F. Sheng, Y. Zhou, X. Yang, Y. Li, C. Feng, and Z.-A. Xu, *Helicity-Protected Ultrahigh Mobility Weyl Fermions in NbP*, Phys. Rev. B **93**, 121112 (2016). DOI: 10.1103/PhysRevB.93.121112

[9] Q. Li, D. E. Kharzeev, C. Zhang, Y. Huang, I. Pletikosić, A. V. Fedorov, R. D. Zhong, J. A. Schneeloch, G. D. Gu, and T. Valla, *Chiral Magnetic Effect in $ZrTe_5$*, Nat. Phys. **12**, 550 (2016). DOI: https://doi.org/10.1038/nphys3648

[10] O. Pavlosiuk, A. Jezierski, D. Kaczorowski, and P. Wiśniewski, *Magnetotransport Signatures of Chiral Magnetic Anomaly in the Half-Heusler Phase ScPtBi*, Phys. Rev. B **103**, 205127 (2021). DOI: 10.1103/PhysRevB.103.205127

[11] P.J.W. Moll, A.C. Potter, N.L. Nair, B.J. Ramshaw, K.A. Modic, S. Riggs, B.Z., N.J. Ghimire, E.D. Bauer, R. Kealhofer, F. Ronning and J.G. Analytis, *Magnetic torque anomaly in the quantum limit of Weyl semimetals*, Nat. Commun. **7**, 12492 (2016). DOI: https://doi.org/10.1038/ncomms12492

[12] J. Klotz, S. C. Wu, C. Shekhar, Y. Sun, M. Schmidt, M. Nicklas, M. Baenitz, M. Uhlarz, J. Wosnitza, C. Felser, and B. Yan, *Quantum Oscillations and the Fermi Surface Topology of the Weyl Semimetal NbP*, Phys. Rev. B **93**, 121105 (2016). DOI: 10.1103/PhysRevB.93.121105

[13] S.-M. Huang, S.-Y. Xu, I. Belopolski, C.-C. Lee, G. Chang, B. Wang, N. Alidoust, G. Bian, M. Neupane, C. Zhang, S. Jia, A. Bansil, H.L. and M.Z. Hasan, *A Weyl Fermion semimetal with surface Fermi arcs in the transition metal monopnictide TaAs class*, Nat. Commun **6**, 7373 (2015). DOI: https://doi.org/10.1038/ncomms8373

[14] H. Weng, C. Fang, Z. Fang, B. Andrei Bernevig, and X. Dai, *Weyl Semimetal Phase in Noncentrosymmetric Transition-Metal Monophosphides*, Phys. Rev. X **5**, 011029 (2015). DOI: 10.1103/PhysRevX.5.011029

[15] C. C. Lee, S. Y. Xu, S. M. Huang, D. S. Sanchez, I. Belopolski, G. Chang, G. Bian, N. Alidoust, H. Zheng, M. Neupane, B. Wang, A. Bansil, M. Z. Hasan, and H. Lin, *Fermi Surface Interconnectivity and Topology in Weyl Fermion Semimetals TaAs, TaP, NbAs, and NbP*, Phys. Rev. B **92**, 235104 (2015). DOI: 10.1103/PhysRevB.92.235104

[16] U. Stockert, R. D. dos Reis, M. O. Ajeesh, S. J. Watzman, M. Schmidt, C. Shekhar, J. P. Heremans, C. Felser, M. Baenitz, and M. Nicklas, *Thermopower and Thermal Conductivity in the Weyl Semimetal NbP*, J. Phys. Condens. Matter **29**, 325701 (2017). DOI: https://doi.org/10.1088/1361-648X/aa7a3b

[17] J. M. Ziman, *Electrons and Phonons* (Clarendon Press, Oxford, 1960).

[18] K. S. Kim, *Role of Axion Electrodynamics in a Weyl Metal: Violation of Wiedemann-Franz Law*, Phys. Rev. B **90**, 121108(2014). DOI: 10.1103/PhysRevB.90.121108

[19] G. Sharma, P. Goswami, and S. Tewari, *Nernst and Magnetothermal Conductivity in a Lattice Model of Weyl Fermions*, Phys. Rev. B **93**, 035116 (2016). DOI: 10.1103/PhysRevB.93.035116

[20] J. J. Heremans, J-P. Issit, A. A. M. Rashidf, and G. A. Saunders, *Electrical and Thermal Transport Properties of Arsenic*, J. Phys. C: Solid State Phys. **10**, 4511 (1977). DOI: https://doi.org/10.1088/0022-3719/10/22/020

[21] A. Jaoui, B. Fauqué, and K. Behnia, *Thermal Resistivity and Hydrodynamics of the Degenerate Electron Fluid in Antimony*, Nat. Commun. **12**, 195 (2021). DOI: https://doi.org/10.1038/s41467-020-20420-9

[22] J. Crossno, J. K. Shi, K. Wang, X. Liu, A. Harzheim, A. Lucas, S. Sachdev, P. Kim, T. Taniguchi, K. Watanabe, T. A. Ohki, and K. C. Fong, *Observation of the Dirac Fluid and the Breakdown of the Wiedemann-Franz Law in Graphene*, Science. **351**, 1058 (2016). DOI: 10.1126/science.aad0343

[23] C. Shekhar, A. K. Nayak, Y. Sun, M. Schmidt, M. Nicklas, I. Leermakers, U. Zeitler, Y. Skourski, J. Wosnitza, Z. Liu, Y. Chen, W. Schnelle, H. Borrmann, Y. Grin, C. Felser, and B. Yan, *Extremely Large Magnetoresistance and Ultrahigh Mobility in the Topological Weyl Semimetal Candidate NbP*, Nat. Phys. **11**, 645 (2015). DOI: 10.1038/NPHYS3372

[24] M. Zarenia, A. Principi, and G. Vignale, *Thermal Transport in Compensated Semimetals: Effect of Electron-Electron Scattering on Lorenz Ratio*, Phys. Rev. B **102**, 214304 (2020). DOI: 10.1103/PhysRevB.102.214304

[25] C. Strohm, G. L. J. A. Rikken, and P. Wyder, *Phenomenological Evidence for the Phonon Hall Effect*, Phys. Rev. Lett. **95**, 155901 (2005). DOI: https://doi.org/10.1103/PhysRevLett.95.155901

[26] G. Grissonnanche, S. Thériault, A. Gourgout, M. Boulanger, E. Lefrançois, A. Ataei, F. Laliberté, M. Dion, J. Zhou, S. Pyon, T. Takayama, H. Takagi, N. Doiron-Leyraud, and L. Taillefer, *Chiral Phonons in the Pseudogap Phase of Cuprates*, Nat. Phys. **16**, 1108 (2020). DOI: https://doi.org/10.1038/s41567-020-0965-y

[27] X. Li, B. Fauqué, Z. Zhu, and K. Behnia, *Phonon Thermal Hall Effect in Strontium Titanate*, Phys. Rev. Lett. **124**, 105901 (2020). DOI: https://doi.org/10.1103/PhysRevLett.124.105901

[28] A. Jaoui, A. Gourgout, G. Seyfarth, A. Subedi, T. Lorenz, B. Benoit Fauqué, and K. Behnia, *Formation of an Electron-Phonon Bi-Fluid in Bulk Antimony*, 2021. arXiv:2105.08408v2

[29] J. M. Ziman, *The Effect of Free Electrons on Lattice Conduction*, Phil. Mag. **1**, 191 (1956). DOI: 10.1080/14786435608238092

[30] J. M. Ziman, *CORRIGENDUM: The Effect of Free Electrons on Lattice Conduction*, Phil. Mag. **2**, 292 (1957). DOI: 10.1080/14786435708243818

[31] S.J. Watzman, T.M. McCormick, C. Shekhar, S.-C. Wu, Y. Sun, A. Prakash, C. Felser, N. Trivedi, and J.P. Heremans, *Dirac dispersion generates unusually large Nernst effect inWeyl semimetals*, Phys. Rev. B **97**, 161404(R). DOI: https://doi.org/10.1103/PhysRevB.97.161404

[32] F.J. Webb, K.R. Wilkinson and J. Wilks, *The thermal conductivity of solid helium*, Proc. Roy. Soc. A, **214**, 546 (1952). DOI: https://doi.org/10.1098/rspa.1952.0188

[33] C. Schindler, D. Gorbunov, S. Zherlitsyn, S. Galeski, M. Schmidt, J. Wosnitza, and J. Gooth, *Strong Anisotropy of the Electron-Phonon Interaction in NbP Probed by Magnetoacoustic Quantum Oscillations*, Phys. Rev. B **102**, 165156 (2020). DOI: 10.1103/PhysRevB.102.165156

[34] K. C. MacDonald, *Thermoelectricity: An Introduction to the Principles* (Courier Corporation, 2013).

**SUPPLEMENTAL MATERIAL:**

**Severe violation of the Wiedemann-Franz law in quantum oscillations of NbP**

Pardeep Kumar Tanwar [1], Md Shahin Alam [1], Dariusz Kaczorowski [2] and Marcin Matusiak [1,2,*]

1. International Research Centre MagTop, Institute of Physics, Polish Academy of Sciences, Aleja Lotnikow 32/46, PL-02668 Warsaw, Poland
2. Institute of Low Temperature and Structure Research, Polish Academy of Sciences, ul. Okólna 2, 50-422 Wrocław, Poland

**A**: Longitudinal resistivity

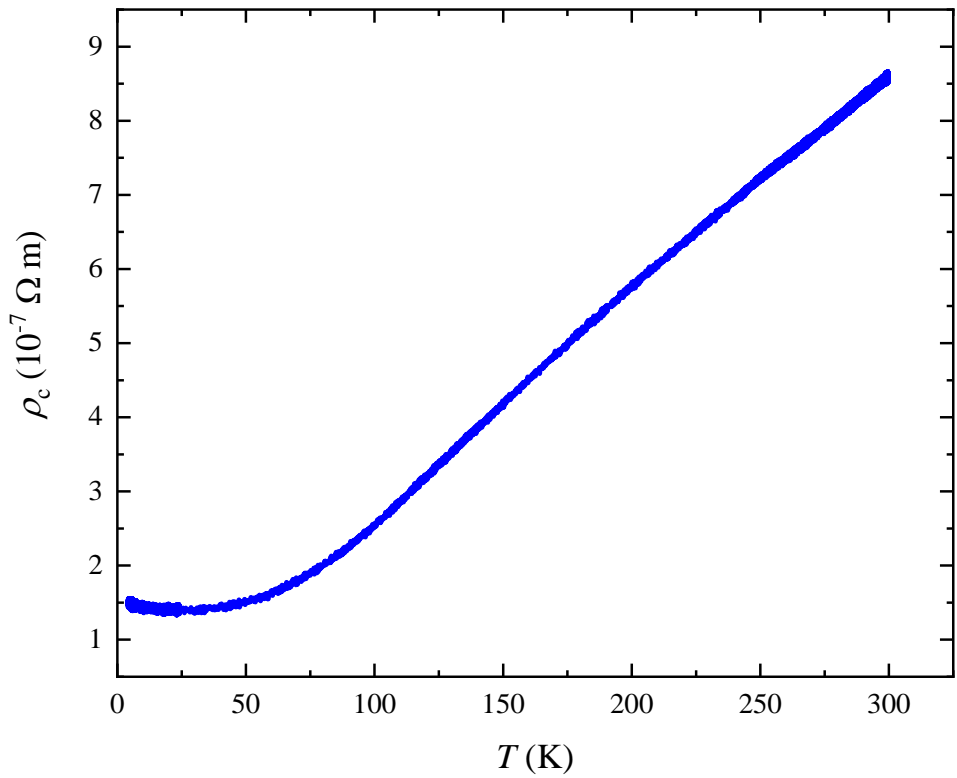

**Figure S1.** Temperature dependence of the resistivity of NbP measured for electric current applied along *c* direction.

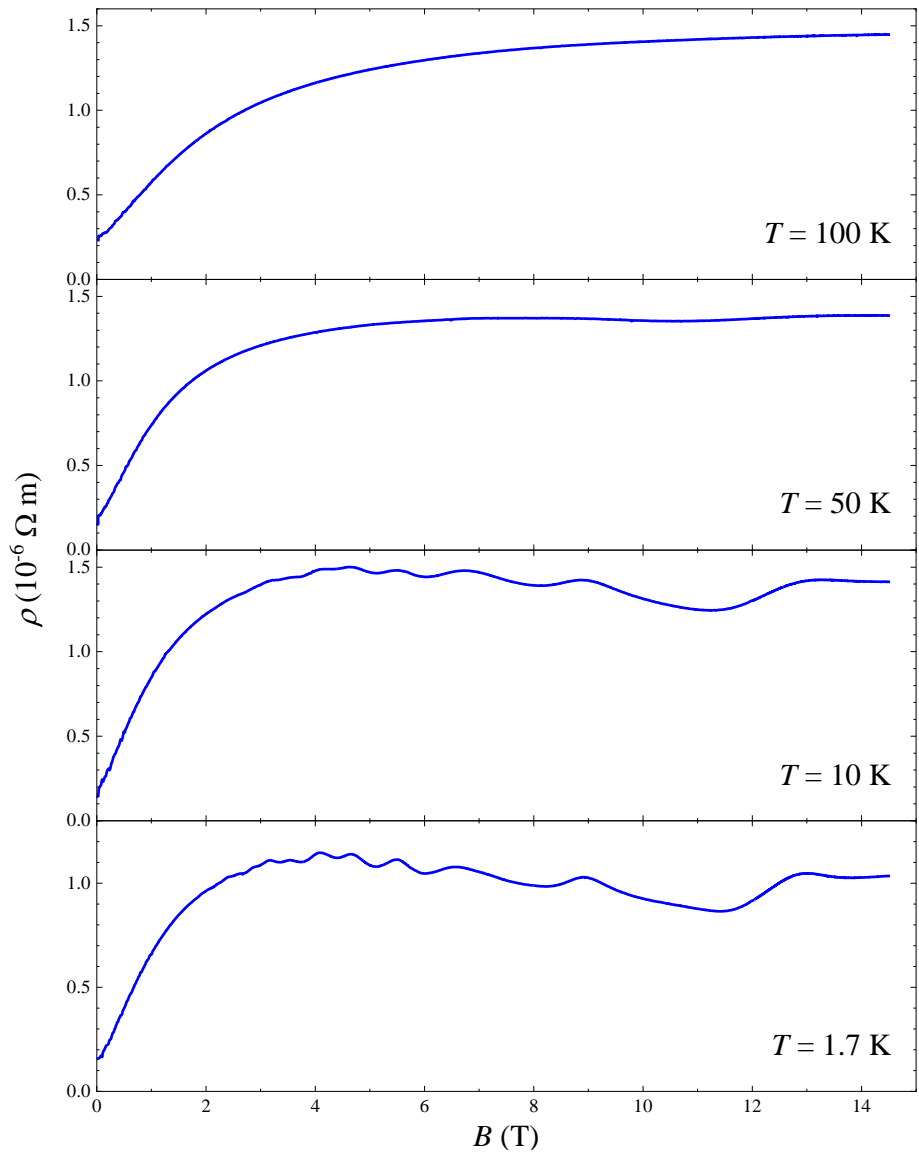

**Figure S2.** Magnetic field dependence of the resistivity of NbP measured for ***B*** ∥ ***c*** ∥ ***J*** at *T* = 1.7, 10, 50 and 100 K (panels from bottom to top, respectively).

**B**: Shubnikov de Haas effect

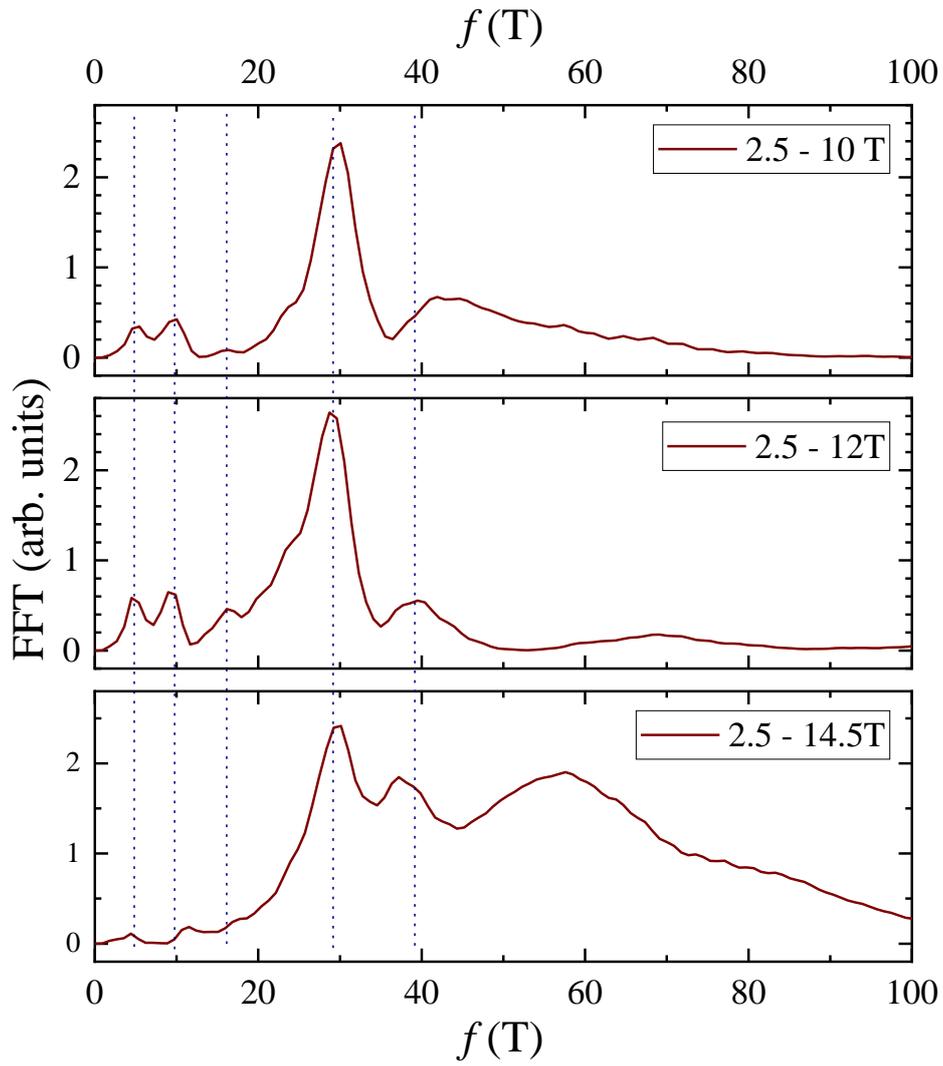

**Figure S3.** Fast Fourier transform of the oscillatory component of $\rho(B)$ at $T = 1.7$ K for magnetic field window set at: 2.5 – 14.5 T, 2.5 – 12 T, 2.5 – 10 T (from the bottom up). In all cases, the signal is dominated by the frequency of ~ 30 T.

**C**: Field dependence of the thermal conductivity

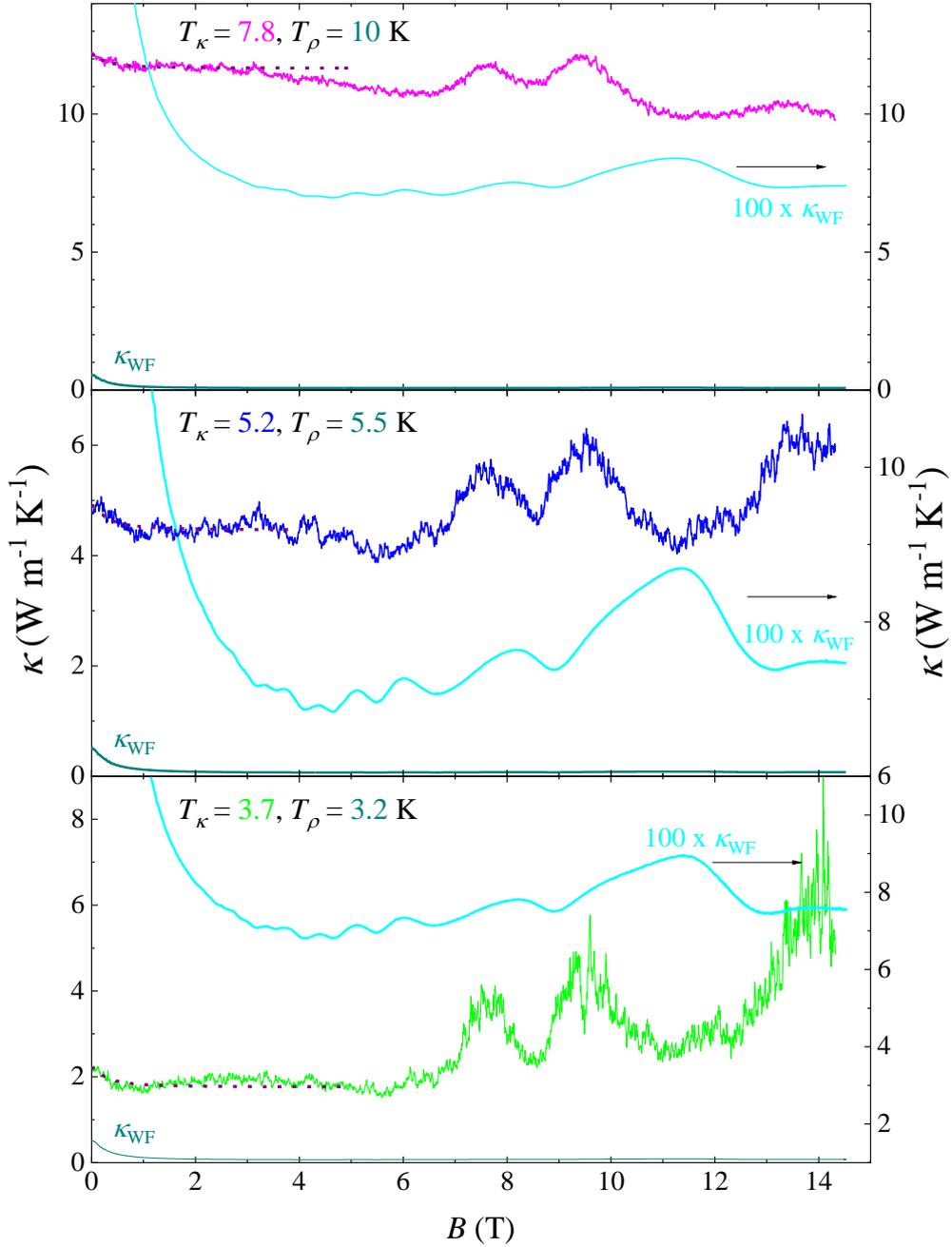

**Figure S4.** Magnetic field dependence of the thermal conductivity of NbP at $T_\kappa$ = 3.7, 5.2, 7.8 K (panels from bottom to top, respectively) compared to the electronic thermal conductivity calculated from the Wiedemann – Franz law at $T_\rho$ = 3.2, 5.5, 10 K (panels from bottom to top, respectively). The dashed line is part of the $\kappa_{WF}(B)$ dependence shifted vertically to match $\kappa_{WF}$ and $\kappa$ values in the zero field limit. The cyan lines show $\kappa_{WF}(B)$ multiplied by 100 (right axes).

**D**: Hall effect

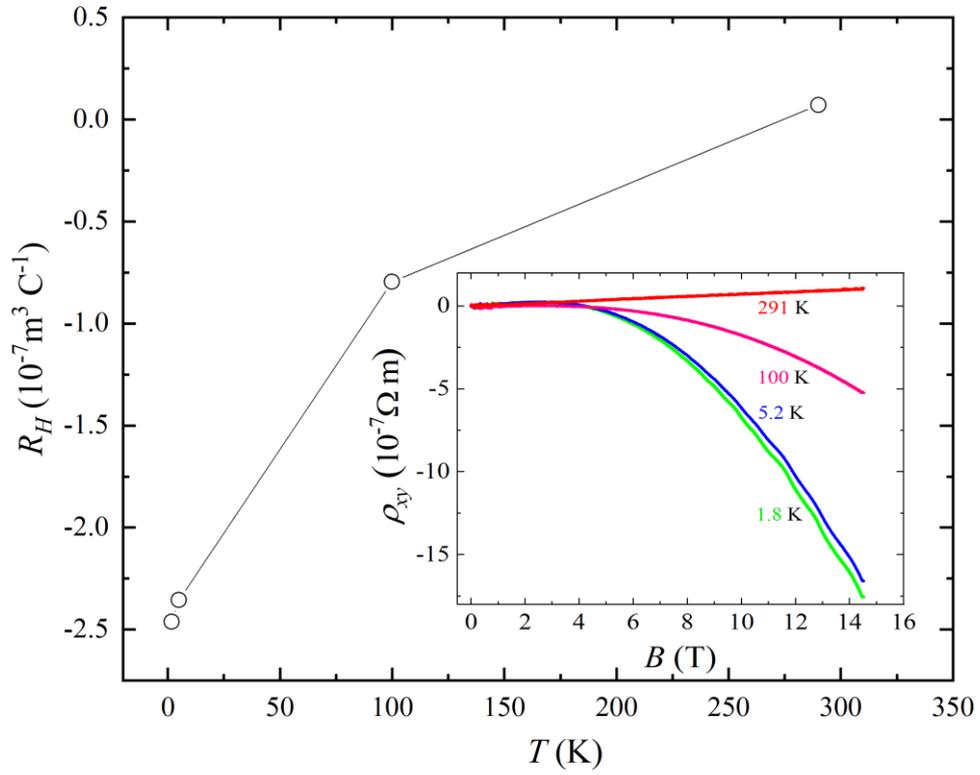

**Figure S5.** Hall coefficient, slope of the magnetic field dependence of the Hall resistivity $\rho_{xy}(B)$ in the high magnetic field regime (10-14.5 T). Inset shows $\rho_{xy}(B)$ measured for $\mathbf{J} \parallel \mathbf{c}$ and the magnetic field oriented parallel to [100] direction at $T_\kappa$ = 1.8 K, 5K, 100K, 291 K (from the top down).

E: Quantitative analysis of the chiral zero sound (CZS) effect

Magnetic frequency:

$$\omega_B = v_F \sqrt{\frac{eB}{\hbar}} \quad (1)$$

Where $v_F = \sqrt{v_x v_y v_z}$ is the mean Fermi velocity of W2 type Weyl points. The Fermi and mean Fermi velocity of W2 type Weyl points are $v_x = v_y \approx 2.1 \times 10^5$ m/s and $v_z \approx 3.8 \times 10^5$ m/s and $v_F = \sqrt{v_x v_y v_z} \approx 2.5 \times 10^5$ m/s [1].

The oscillating part of compressibility:

$$\frac{\beta^{(1)}}{\beta^{(0)}} = \hbar \frac{\omega_B}{\mu} \frac{\exp\left(-\pi \frac{\mu}{\hbar^2 \omega_B^2 \tau_0}\right)}{\sinh\left(2\pi^2 \frac{\mu k_B T}{\hbar^2 \omega_B^2}\right)} \cos\left(-\pi \frac{\mu^2}{\hbar^2 \omega_B^2} - \frac{\pi}{4}\right) \quad (2)$$

Where $\sinh(x) = \frac{(e^x - e^{-x})}{2x} = \frac{\sinh(x)}{x}$ and electronic relaxation time $\tau_0 = \frac{6\pi^2 \sigma^{elec} \hbar}{N_{WP}(e^2 k_F^2 v_F)} \approx 5.3 \times 10^{-13}$ s, estimated from Drude model for Weyl semimetal. The Fermi wavevector $\kappa_F \approx 0.031 \times 10^{10}$ m$^{-1}$ for frequency equal to $\approx 32\,T$ [2], and total number of Weyl points $N_{WP} = 24$.

Contribution of chiral zero sound ($\kappa_{CZS}(B)$) to total thermal conductivity ($k(B)$):

$$\kappa_{CZS}(B) = \tau_s k_B \Lambda^3 v_F^2 N_{czs} \left(\frac{(\hbar \omega_B)^2}{4\pi^2 \mu^2}\left(1 - \frac{\beta^{(1)}}{\beta^{(0)}}\right)\right)^2 F\left(\frac{\hbar v_F \Lambda}{k_B T} \frac{(\hbar \omega_B)^2}{4\pi^2 \mu^2}\left(1 - \frac{\beta^{(1)}}{\beta^{(0)}}\right)\right) \quad (3)$$

Where $F(a) = \frac{1}{a^3} \int_{-a}^{a} \frac{x^2(a^2 - x^2)}{32\pi^2 \sinh^2\left(\frac{x}{2}\right)} dx$, the cutoff wavevector of CZS mode $\Lambda = 2\kappa_F \left(\sqrt{\frac{v_x v_y}{v_z}}\right)$ and $N_{czs} = 18$, Weyl points which contribute to CZS. A factor ($\sqrt{\frac{v_x v_y}{v_z}}$) is multiplied due to anisotropy of Fermi surface along z.

**Velocity of chiral zero sound**:

$$c(B) = \frac{eB}{4\pi^2 \beta(B)} \sqrt{(\xi_0^2 - \xi_1^2)} \tag{4}$$

Where $\beta(B) = \frac{\mu^2}{2\pi^2 v_F^3}$

Factor $\sqrt{(\xi_0^2 - \xi_1^2)}$ was calculated from [3], since, NbP and TaAs have similar electronic structure.

$$c(B) \approx 1.4 \times 10^4 \text{ m/s for } B=14.5 \text{ T}$$

**Debye temperature of chiral zero sound:**

$$\Theta_{CZS} = \hbar c(B) \Lambda / k_B \tag{5}$$

$$\Theta_{CZS} \approx 55.3 \, K \text{ for } B = 14.5 \text{ T}$$

**Relaxation time for chiral zero sound:**

$$\bar{\kappa} = \frac{\kappa_{CZS}(B)}{\tau_s k_B \Lambda^3 v_F^2 N_{czs}} \tag{6}$$

Where $\bar{\kappa}$ is dimensionless thermal conductivity, used to calculate $\tau_s$ (relaxation time for chiral zero sound). $\bar{\kappa} = 7.8 \times 10^{-7}$ for $B = 13.6$ T at $T = 5.2$ K.

$$\tau_s \approx 1.7 \times 10^{-9} \text{ s}$$

**Chiral limit:**

$$\frac{\tau_0}{\tau_s} \approx 4.5 \times 10^{-4}$$

References:

[1] C. C. Lee, S. Y. Xu, S. M. Huang, D. S. Sanchez, I. Belopolski, G. Chang, G. Bian, N. Alidoust, H. Zheng, M. Neupane, B. Wang, A. Bansil, M. Z. Hasan, and H. Lin, *Fermi Surface Interconnectivity and Topology in Weyl Fermion Semimetals TaAs, TaP, NbAs, and NbP*, Phys. Rev. B **92**, 235104 (2015). DOI: 10.1103/PhysRevB.92.235104

[2] C. Shekhar, A. K. Nayak, Y. Sun, M. Schmidt, M. Nicklas, I. Leermakers, U. Zeitler, Y. Skourski, J. Wosnitza, Z. Liu, Y. Chen, W. Schnelle, H. Borrmann, Y. Grin, C. Felser, and B. Yan, *Extremely Large Magnetoresistance and Ultrahigh Mobility in the Topological Weyl Semimetal Candidate NbP*, Nat. Phys. **11**, 645 (2015). DOI: 10.1038/NPHYS3372

[3] Z. Song and X. Dai, *Hear the Sound of Weyl Fermions*, Phys. Rev. X **9**, 021053 (2019). DOI: https://doi.org/10.1103/PhysRevX.9.021053